# SOLITON ANALYSIS OF THE ELECTRO-OPTICAL RESPONSE OF BLUE BRONZE

J.W. Brill*

Department of Physics and Astronomy, University of Kentucky,

Lexington, KY 40506-0055, USA

ABSTRACT

In recent measurements on the charge-density-wave (CDW) conductor blue bronze ($K_{0.3}MoO_3$), the electro-transmittance and electro-reflectance spectra were searched for intragap states that could be associated with solitons created by injection of electrons into the CDW at the current contacts [Eur. Phys. J. B **16**, 295 (2000); *ibid* **35**, 233 (2003)]. In this work, we adapt the model of soliton absorption in dimerized polyacetylene to the blue bronze results, to obtain the (order of magnitude) estimate that current induced solitons occur on less than ~ 10% of the conducting chains. We discuss the implications of these results on models of soliton lifetimes and motion of CDW phase dislocations.

* jwbrill@uky.edu; Tel: 1-859-257-4670; FAX: 1-859-323-2846

The phenomenon of collective transport by sliding charge-density-waves (CDWs) in quasi-one dimensional conductors has been known for almost three decades [1,2], but our understanding of the details of "current conversion", i.e. how electrons enter and leave the CDW at current contacts, has remained incomplete. Macroscopically, conversion is believed to occur through CDW phase-slip, occurring through the climb of CDW phase dislocations across the sample cross-section [3]. In a series of elegant transport measurements on the semimetallic CDW conductor, NbSe$_3$ [1], the Cornell group has shown how phase-slip is driven by longitudinal strains in the CDW [4]. The strain profiles they deduced have been qualitatively verified with x-ray diffraction [5] and infrared electro-optic measurements [6]. The latter measurements, the subject of this paper, were done on semiconducting $K_{0.3}MoO_3$ (blue bronze) [7].

There has been less progress in our attempts to observe and measure the electronic defect states constituting the phase dislocations. In the prevailing model of Brazovskii and Matveenko [8], an electron falls into a midgap state as it enters the conducting chain. This midgap state is associated with a "π-soliton", i.e. a localized half-wavelength deformation of the CDW [9]. Two π-solitons quickly (e.g. within a phonon period) condense into a 2π-soliton, with energy states distributed within ~ $k_BT_c$ of the edge of the CDW gap, where $T_c$ (= 180 K for blue bronze [7]) is the CDW transition temperature [8,10]. 2π-solitons on neighboring chains coalesce into the phase dislocation. However, although excitations of solitons present as "growth defects" of the CDW *may* have been observed in the IR spectrum of the CDW material TaS$_3$ [11], soliton states have not yet been observed to accompany CDW sliding.

In the first electro-optic experiments, it was found that, for light polarized perpendicular to the conducting chains, the infrared transmittance ($\tau$) of blue bronze varied with position when a voltage larger than the CDW depinning threshold was applied, and the relative change in transmission, ($\Delta\tau/\tau_\perp$) was assumed to be proportional to the local CDW phase gradient [6]. By comparing $\Delta\tau/\tau_\perp$ spectra for different voltages and positions in the sample we hoped to observe absorption lines that could be associated with excitation of the solitons (i.e. soliton-to-conduction band or valence band-to-soliton transitions) created by the injected current [12], in analogy to doping induced soliton absorption in conducting polymers [13,14]. However, the normalized $\Delta\tau/\tau_\perp$ spectrum was found to be independent of voltage and position (to precision $\delta\Delta\tau/\tau_\perp < 4 \times 10^{-5}$) even at positions adjacent (within the phase-slip length $\Lambda \sim 0.1$ mm) to current contacts and under conditions of "rapid" current conversion (e.g. washboard or "narrow-band-noise" frequencies [2] $f_{NBN} = v_{CDW}/\lambda > 30$ kHz, where $\lambda$ and $v_{CDW}$ are the wavelength and average velocity of the CDW) [12]. This result, that no new absorption lines were observed near the contacts, was interpreted as meaning that $2\pi$-solitons (i.e. phase slip centers) occurred on less than a few percent of the conducting chains at any time [12].

However, we also pointed out that soliton absorption might be preferentially polarized along the conducting chains, for which the strong background absorption makes transmission measurements impossible. We therefore studied the parallel polarized electro-reflectance ($\Delta R/R_{//}$) spectra [15], and found that it was also independent of position and voltage (to precision $\delta\Delta R/R_{//} < 2.6 \times 10^{-6}$) implying that solitons occurred on less than 15% of the chains.

The above estimates of the upper limit of soliton density were made using two somewhat unphysical simplifications: i) We assumed that, although the absorption would vary with energy (i.e. there would be an absorption peak of width $\Gamma \sim k_B T_c$), the average absorption cross-section within the peak would simply be the geometric cross-section $\sigma_0 \sim \xi_{//}\xi_{\perp}$ of the soliton. Here $\xi_{//} \sim 20$ Å and $\xi_{\perp} \sim 5$ Å $\sim$ the interchain spacing are the CDW amplitude coherence lengths along and transverse to the chains [16]. ii) We did not use the fact that we searched for the soliton absorption spectroscopically, i.e. by varying the wavelength of the light source "continuously", but instead our estimate was appropriate for attempting to observe the extra absorption with a source at one wavelength (or several discretely separated wavelengths). Hence our estimates depended on the linewidth of our light source. In this paper, we make a more physical calculation of the expected soliton absorption; the resulting estimates for the upper limits of soliton density are slightly higher than our previous ones.

As far as we know, there have been no calculations of the soliton-band absorption for an incommensurate CDW (i.e. those in materials for which the CDW can be easily depinned [2]). However, there have been several calculations for the case of polyacetylene, for which the CDW corresponds to a dimerization of the lattice [14]. However, since the polyacetylene results were obtained in a continuum model, the details of the atomic displacements should not be important. The absorption coefficient at photon energy $\nu$, for light polarized along the chains, was calculated to be [14]:

$$\Delta\alpha(\nu) = N(e^2/\hbar c \varepsilon_1^{1/2})\pi\xi_{//}^2 x^{-1/2}\text{sech}^2(x), \quad x \equiv (\nu/\nu_0)^2 - 1, \quad (1)$$

where N is the soliton density, $\nu_0$ is the soliton excitation energy, and $\varepsilon \equiv \varepsilon_1 + \iota\varepsilon_2$ is the dielectric constant. (For polyacetylene, $\nu_0$ is half the gap [14], but we keep it as arbitrary for our case.) One of the powers of the coherence length $\xi_\parallel$ comes from the geometric size of the delocalized soliton [13] and the other from the induced dipole moment in the optical field. We therefore replace $\xi_\parallel^2$ with $\xi_\parallel \xi_\perp$ (i.e. $\sigma_0$) for transversely polarized light.

The $x^{-1/2}$ behavior in Eqtn. (1) reflects the quasi-one-dimensional band structure. As discussed in [17], this divergence at $\nu_0$ is generally not observed, but a more symmetric peak in $\alpha$ is observed instead (e.g. see [13,14]). Hence, we replace the behavior of Eqtn. (1) with a peak of width $\Gamma$ with the same integrated intensity:

$$\int \Delta\alpha(\nu) d\nu \sim 0.6\, N(e^2/\hbar c \varepsilon_1^{1/2})\pi \xi^2 \nu_0, \quad (2)$$

where $\xi^2 = \xi_\parallel^2$ for parallel polarization and $\xi^2 = \xi_\parallel \xi_\perp$ for perpendicular. For concreteness, we take a Lorentzian peak in the dielectric constant:

$$\varepsilon = \varepsilon_\infty + M\Gamma\nu_0 / [(\nu_0^2 - \nu^2) - i\Gamma\nu], \quad (3)$$

with $\varepsilon_{\infty\parallel} = 70 + 65i$ and $\varepsilon_{\infty\perp} = 12 + 6i$, taken to match the measured spectra of blue bronze [18] near 1000 cm$^{-1}$. The integral of the change in absorption coefficient, $\Delta\alpha = (4\pi\nu)\Delta\text{Im}(\varepsilon^{1/2})$, for $\nu$ expressed in wavenumbers, due to the peak is then

$$\int \Delta\alpha(\nu) d\nu \sim 2^{3/2} \pi \nu_0 \Gamma M / \{[(\varepsilon_{10}^2 + \varepsilon_{20}^2)^{1/2} - \varepsilon_{10}] [(\varepsilon_{10}/\varepsilon_{20})^2 + 1]\}^{1/2}, \quad (4)$$

so that, comparing with Eqtn. (2), the oscillator strength is

$$M \sim 0.0015\, (N\xi^2/\Gamma) \{[(\varepsilon_{10}^2 + \varepsilon_{20}^2)^{1/2} - \varepsilon_{10}] [(\varepsilon_{10}/\varepsilon_{20})^2 + 1] / \varepsilon_{10}\}^{1/2} \quad (5)$$

Figure (1) shows the resulting relative changes in parallel polarized reflectance, $R = |[\varepsilon^{1/2}-1]/[\varepsilon^{1/2}+1]|^2$, and perpendicularly polarized transmission, $\tau \sim (1-R)^2 \exp(-\alpha d)$ (for a sample with $\alpha d > 1$), where $d = 3.5$ μm is the thickness of the crystal [12], assuming that there is one soliton/conducting chain in the $\Lambda \sim 0.1$ mm phase slip region (i.e. $N \sim 10^{16}$ cm$^{-3}$). For these calculations, we took $\Gamma = k_B T_c = 125$ cm$^{-1}$ [7] and $\nu_0 = 1100$ cm$^{-1}$, near the edge of the gap [12]. The resulting peak values are $\Delta\tau/\tau_\perp \sim 6 \times 10^{-4}$ and $\Delta R/R_\parallel \sim 9 \times$

$10^{-6}$. The experimental upper limits for the variation of the electro-optic signals, $\delta\Delta\tau/\tau_\perp <$ 4 x $10^{-5}$ [12] and $\delta\Delta R/R_{//} <$ 2.6 x $10^{-6}$ [15], therefore would imply that solitons occur on less than 7% and 30%, respectively, of the conducting chains at any time. Given the many approximations made, these of course should just be considered order of magnitude estimates, but our results certainly imply that there is less than one soliton (or, more precisely, intragap state) per chain.

The rate of injection of $2\pi$-solitons into each conducting chain is given by the narrow-band-noise frequency [2,12], so their density in the phase-slip region is given by $N = f_{NBN} T_0/ (\Lambda\Omega)$, where $\Omega$ is the area/chain and $T_0$ is the "lifetime" for which the intragap state can be optically excited. Hence, there will be one state/chain if $T_0 = 1/f_{NBN}$, as discussed in Reference [12,15]. For example, in [15] we assumed that the soliton moves off its original chain when a second $2\pi$-soliton is created, i.e. when the CDW advances one wavelength. If the intragap states then became optically inert, e.g. as the soliton moved into a phase dislocation, one would have $T_0 = 1/f_{NBN}$. However, if the intrgap states remain optically excitable within the dislocation as it moves through the crystal of thickness d, then $T_0 \sim d/(\Omega^{1/2} f_{NBN})$, and the density of the intragap states would be orders of magnitude larger.

Alternatively, Ong and Maki [20] assumed that each phase dislocation lines swept through the entire sample cross-section in the narrow-band-noise period. In that case, $T_0 \sim 1/f_{NBN}$ if the intragap states are excitable for this whole time, but $T_0 \sim \Omega^{1/2}/(df_{NBN})$ if they can only be excited before the soliton moves off its orginal chain and coalesces into a phase dislocation.

In summary, we have used the soliton absorption model of polyacetylene [14,17] to reanalyze our electro-transmission [12] and electro-reflectance measurements [15] in blue bronze. From the transmission results, we find that solitons created by charge injection into the CDW occur on less than ~ 10% of the conducting chains at any instant (with the limit a few times higher from the reflectance measurement). While this should only be considered an order of magnitude estimate, it certainly implies that the optically excitable lifetime of the soliton is less than the narrow-band-noise period. This in turn implies that the intragap state becomes optically inert once the solitons on neighboring chains coalesce into CDW phase dislocations.

Helpful discussions with R.C. Rai and T. Troland are gratefully acknowledged. This research was supported by the U.S. National Science Foundation, Grant # DMR-0100572.

**REFERENCES**

[1] P. Monceau, N.P. Ong, A.M. Portis, A. Meerschaut, and J. Rouxel, Phys. Rev. Lett. **37**, 602 (1976).

[2] G. Grüner, Rev. Mod. Phys. **60**, 1129 (1988).

[3] J.C. Gill, Solid State Commun. **44**, 1041 (1982); J.C. Gill, J. Phys. IV France **9**, Pr10-97 (1999); N. Kirova and S. Brazovskii, J. Phys. IV France **12**, Pr9-173 (2002).

[4] T.L. Adelman, M.C. de Lind van Wijngaarden, S.V. Zaitwev-Zotov, D. DiCarlo, and R.E. Thorne, Phys. Rev. B **53**, 1833 (1996); S.G. Lemay, .C. de Lind van Wijngaarden, T.L. Adelman, and R.E. Thorne, Phys. Rev. B **57**, 12781 (1998).

[5] S. Brazovskii, N. Kirova, H. Requardt, F. Ya. Nad, P. Monceau, R. Currat, J.E. Lorenzo, G. Grübel, and Ch. Vettier, Phys. Rev. B **61**, 10640 (2000).

[6] M.E. Itkis, B.M. Emerling, and J.W. Brill, Phys. Rev. B **52**, R11545 (1995).

[7] C. Schlenker, C. Filippini, J. Marcus, J. Dumas, J.P. Pouget, and S. Kagoshima, J. Phys. (France) **44**, C3-1757 (1983).

[8] S. Brazovskii and S. Matveenko, J. Phys. I France **1**, 269, 1173 (1991).

[9] S. Brazovskii, Sov Phys JETP **51**, 342 (1980); S. N. Artemenko, JETP Lett. **63**, 56 (1995).

[10] S. Brazovskii, private communication.

[11] F. Ya. Nad' and M.E. Itkis, JETP Lett. **63**, 262 (1996); G. Minton and J.W. Brill, Solid State Commun. **65**, 1069 (1988).

[12] B.M. Emerling, M.E. Itkis, and J.W. Brill, Eur. Phys. J. B **16**, 295 (2000).

[13] H. Suzuki, M. Ozaki, S. Etemad, A.J. Heeger, and A.G. MacDiarmid, Phys. Rev. Lett. **45**, 1209 (1980).

[14] A.J. Heeger, S. Kivelson, J.R. Schrieffer, and W.-P. Su, Rev. Mod. Phys. **60**, 781 (1988).

[15] R.C. Rai, V.A. Bondarenko, and J.W. Brill, Eur. Phys. J. B **35**, 233 (2003).

[16] S. Girault, A.H. Moudden, and J.P. Pouget, Phys. Rev. B **39**, 4430 1989).

[17] S. Kivelson, Ting-Kuo Lee, Y.R. Lin-Liu, Ingo Peschel, and Lu Yu, Phys. Rev. B **25**, 4173 (1982).

[18] G. Travaglini, P. Wachter, J. Marcus, and C. Schlenker, Solid State Commun. **37**, 599 (1981); S. Jandl, M. Banville, C. Pepin, J. Marcus, and C. Schlenker, Phys. Rev. B **40**, 12487 (1989).

[19] For example, the absorption coefficient of polyacetylene calculated from Eqtn. (1) underestimates its observed value by a few times [17].

[20] N.P. Ong and K. Maki, Phys. Rev. B **32**, 6582 (1985).

**FIGURE CAPTION**

Figure 1. Calculated spectra of the relative changes in reflectance (parallel polarization) and transmission (perpendicular polarization) for one soliton/chain, using the parameters discussed in the text. (Note the different orders of magnitude for the two spectra and that the $\Delta\tau/\tau$ spectrum is inverted.)

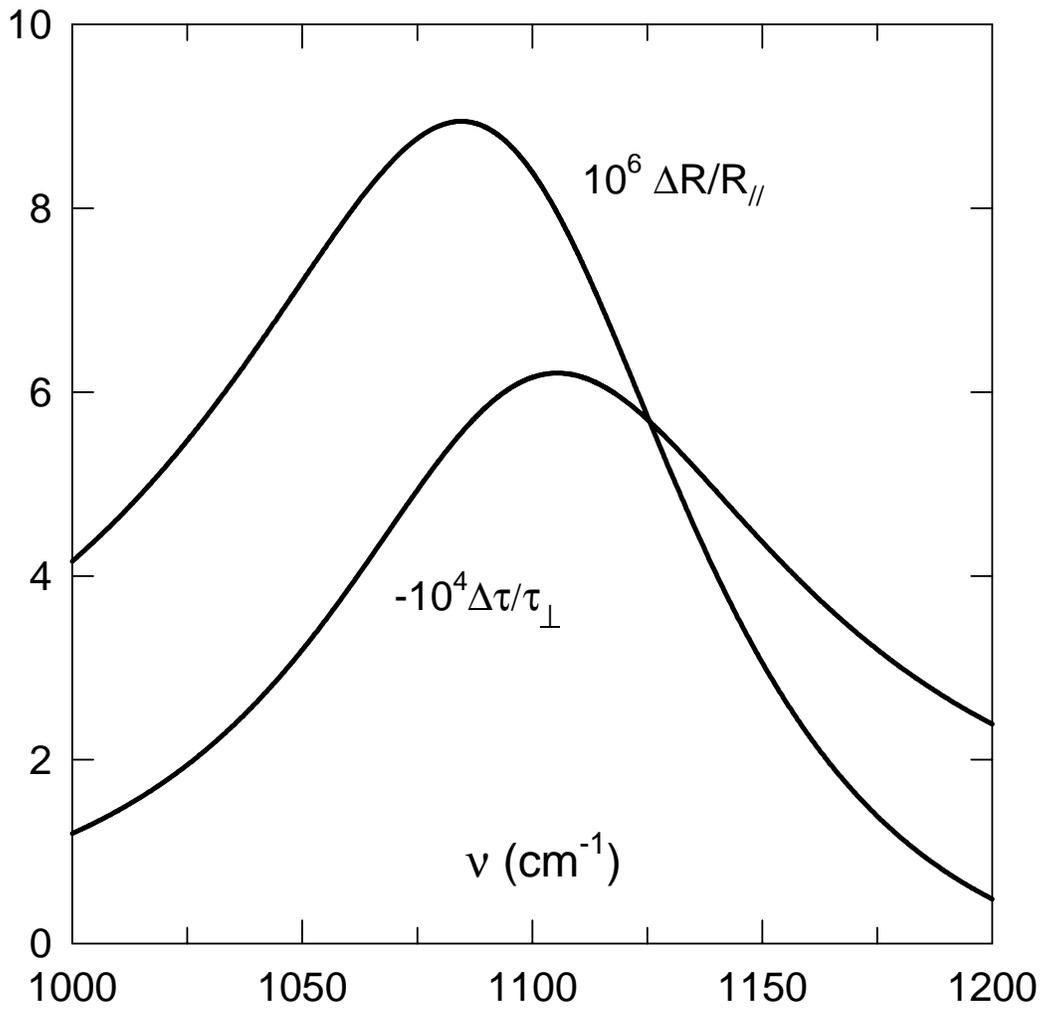